\documentclass[12pt]{article}%
\usepackage{amsmath}
\usepackage{amsfonts}
\usepackage{mitpress}
\usepackage{amssymb}
\usepackage{graphicx}%
\providecommand{\U}[1]{\protect\rule{.1in}{.1in}}
\newtheorem{theorem}{Theorem}

\newtheorem{lemma}[theorem]{Lemma}

\newtheorem{remark}[theorem]{Remark}

\newdimen\dummy
\dummy=\oddsidemargin
\begin{document}

\title{Minimum divergence estimators, maximum likelihood and exponential families}
\author{Michel Broniatowski\\LSTA Universit\'{e} Pierre et Marie Curie\\5 Place Jussieu, 75005 Paris, France\\e-mail: michel.broniatowski@upmc.fr}
\maketitle

\begin{abstract}
In this note we prove the dual representation formula of the divergence
between two distributions in a parametric model. Resulting estimators for the
divergence as for the parameter are derived.\ These estimators do not make use
of any grouping nor smoothing. It is proved that all differentiable
divergences induce the same estimator of the parameter on any regular
exponential family, which is nothing else but the MLE.


\end{abstract}

\section{\bigskip Introduction}

\subsection{Context and scope of this note}

This note presents a short proof of the duality formula \ for $\varphi-$
divergences defined through differentiable convex functions $\varphi$ in
parametric models and discusses some unexpected phenomenon in the context of
exponential families. First versions of this formula appear in \cite{LV1987} p
33, in \cite{B2004} in the context of the Kullback-Leibler divergence and in
\cite{K2003} in a general form. The paper \cite{BL2006} introduces this form
in the context of minimal $\chi^{2}-$ estimation; a global approach to this
formulation is presented in Broniatowski and K\'{e}ziou (2006)\cite{BK2006}.
Independently Liese and Vajda (2006)\cite{LV2006} have obtained a similar
expression based on a much simpler argument as presented in all the above
mentioned papers (formula (118) in their paper); however the proof of their
result is merely sketched and we have found it useful to present a complete
treatment of this interesting result in the parametric setting, in contrast
with the aforementioned approaches.

The main interest of the resulting expression is that it leads to a wide
variety of estimators, by a plug in method of the empirical measure evaluated
on the current data set; so, for any type of sampling its estimators and
inference procedures, for any $\varphi-$divergence criterion. In the case of
the simple i.i.d. sampling resulting properties of those estimators and
subsequent inferential procedures are studied in \cite{BK2009}.

A striking fact is that all minimum divergence estimators defined through this
dual formula coincide with the MLE in exponential families.\ They henceforth
enjoy strong optimality under the standard exponential models, leading to
estimators different from the MLE\ under different models than the exponential
one.\ Also this result proves that MLE\ 's of parameters of exponential
families are strongly motivated by being generated by the whole continuum of
$\varphi-$divergences.

This note results from joint cooperation with late Igor Vajda.

\subsection{Notation}

Let $\mathcal{P}:=\left\{  P_{\theta},\theta\in\Theta\right\}  $ an
identifiable parametric model on $\mathbb{R}^{d}$ where $\Theta$ is a subset
of $\mathbb{R}^{s}.$ All measures in $\mathcal{P}$ will be assumed to be
measure equivalent sharing therefore the same support. The parameter space
$\Theta$ need not be open in the present setting. It may even happen that the
model includes measures which would not be probability distributions; cases of
interest cover models including mixtures of probability distributions; see
\cite{BK2009}. Let $\varphi$ be a proper closed convex function from
$]-\infty,+\infty\lbrack$ to $[0,+\infty]$ with $\varphi(1)=0$ and such that
its domain $\text{dom}\varphi:=\left\{  x\in\mathbb{R}\text{ such that
}\varphi(x)<\infty\right\}  $ is an interval with endpoints $a_{\varphi
}<1<b_{\varphi}$ (which may be finite or infinite). For two measures
$P_{\alpha}$ and $P_{\theta}$ in $\mathcal{P}$ the $\varphi$-divergence
between $Q$ and $P$ is defined by
\[
\phi(\alpha,\theta):=\int_{\mathcal{X}}\varphi\left(  \frac{dP_{\alpha}%
}{dP_{\theta}}(x)\right)  ~dP_{\theta}(x).
\]
In a broader context, the $\varphi$-divergences were introduced by
\cite{C1963} as \textquotedblleft$f$-divergences\textquotedblright. The basic
property of $\varphi-$ divergences states that when $\varphi$ is strictly
convex on a neighborhood of $x=1$, then
\[
\phi(\alpha,\theta)=0~\text{ if and only if }~\alpha=\theta.
\]
We refer to \cite{LV1987} chapter 1 for a complete study of those properties.
Let us simply quote that in general $\phi(\alpha,\theta)$ and $\phi
(,\theta,\alpha)$are not equal. Hence, $\varphi$-divergences usually are not
distances, but they merely measure some difference between two measures. A
main feature of divergences between distributions of random variables $X$ and
$Y$ is the invariance property with respect to common smooth change of variables.

\subsection{Examples of $\varphi$-divergences}

\noindent The Kullback-Leibler $(KL)$, modified Kullback-Leibler $(KL_{m})$,
$\chi^{2}$, modified $\chi^{2}$ $(\chi_{m}^{2})$, Hellinger $(H)$, and $L_{1}$
divergences are respectively associated to the convex functions $\varphi
(x)=x\log x-x+1$, $\varphi(x)=-\log x+x-1$, $\varphi(x)=\frac{1}{2}{(x-1)}%
^{2}$, $\varphi(x)=\frac{1}{2}{(x-1)}^{2}/x$, $\varphi(x)=2{(\sqrt{x}-1)}^{2}$
and $\varphi(x)=\left\vert x-1\right\vert $. All these divergences except the
$L_{1}$ one, belong to the class of the so called \textquotedblleft power
divergences\textquotedblright\ introduced in \cite{CR1988} (see also
\cite{LV1987} chapter 2), a class which takes its origin from R\'{e}nyi
\cite{R1961}. They are defined through the class of convex functions
\begin{equation}
x\in]0,+\infty\lbrack\mapsto\varphi_{\gamma}(x):=\frac{x^{\gamma}-\gamma
x+\gamma-1}{\gamma(\gamma-1)} \label{gamma convex functions}%
\end{equation}
if $\gamma\in\mathbb{R}\setminus\left\{  0,1\right\}  $, $\varphi
_{0}(x):=-\log x+x-1$ and $\varphi_{1}(x):=x\log x-x+1$. So, the
$KL$-divergence is associated to $\varphi_{1}$, the $KL_{m}$ to $\varphi_{0}$,
the $\chi^{2}$ to $\varphi_{2}$, the $\chi_{m}^{2}$ to $\varphi_{-1}$ and the
Hellinger distance to $\varphi_{1/2}$.\newline

\noindent It may be convenient to extend the definition of the power
divergences in such a way that $\phi(\alpha,\theta)$ may be defined (possibly
infinite) even when $P_{\alpha}$ or $P_{\theta}$ is not a probability
measure.\ This is achieved setting
\begin{equation}
x\in]-\infty,+\infty\lbrack\mapsto\left\{
\begin{array}
[c]{lll}%
\varphi_{\gamma}(x) & \text{ if } & x\in\lbrack0,+\infty\lbrack,\\
+\infty & \text{ if } & x\in]-\infty,0[.
\end{array}
\right.  \label{gamma convex functions sur R}%
\end{equation}
when dom$\varphi=\mathbb{R}^{+}/\left\{  0\right\}  .$ Note that for the
$\chi^{2}$-divergence, the corresponding $\varphi$ function $\phi
_{2}(x):=\frac{1}{2}(x-1)^{2}$ is defined and convex on whole $\mathbb{R}$.
\newline

We will only consider divergences defined through differentiable functions
$\varphi$, which we assume to satisfy

(\textbf{RC})~~~~
\begin{tabular}
[c]{l}%
There exists a positive $\delta$ such that for all $c$ in $\left[
1-\delta,1+\delta\right]  $,\\
we can find numbers $c_{1},c_{2},$ $c_{3}$ such that\\
$\varphi(cx)\leq c_{1}\varphi(x)+c_{2}\left\vert x\right\vert +c_{3}$, for all
real $x$.
\end{tabular}

Condition (\textbf{RC}) holds for all power divergences including $KL$ and
$KL_{m}$ divergences.

\section{\noindent Dual form of the divergence and dual estimators in
parametric models}

Let $\theta$ and $\theta_{T}$ be any parameters in $\Theta.$ We intend to
provide a new expression for $\phi(\theta,\theta_{T}).$

By strict convexity, for all $a$ and $b$ $\ $the domain of $\varphi$ it holds%
\begin{equation}
\varphi(b)\geq\varphi(b)+\varphi^{\prime}(a)(b-a) \label{convevity of phi}%
\end{equation}
with equality if and only if $a=b.$

Denote
\[
\varphi^{\#}\left(  x\right)  :=x\varphi^{\prime}(x)-\varphi(x).
\]

For any $\alpha$ in $\Theta$ denote%
\[
a:=\frac{dP_{\theta}}{dP_{\alpha}}\left(  x\right)  .
\]
Define%
\[
b:=\frac{dP_{\theta}}{dP_{\theta_{T}}}\left(  x\right)  .
\]

Inserting these values in (\ref{convevity of phi}) and integrating with
respect to $P_{\theta_{T}}$ yields%
\[
\phi(\theta,\theta_{T})\geq\int\left[  \varphi^{\prime}\left(  \frac
{dP_{\theta}}{dP_{\alpha}}\right)  dP_{\theta}-\varphi^{\#}\left(
\frac{dP_{\theta}}{dP_{\alpha}}\right)  \right]  dP_{\theta_{T}}.
\]
Assume at present that this entails
\begin{equation}
\phi(\theta,\theta_{T})\geq\int\varphi^{\prime}\left(  \frac{dP_{\theta}%
}{dP_{\alpha}}\right)  dP_{\theta}-\int\varphi^{\#}\left(  \frac{dP_{\theta}%
}{dP_{\alpha}}\right)  dP_{\theta_{T}} \label{ineg}%
\end{equation}
for suitable $\alpha$'s in some set $\mathcal{F}_{\theta}$ included in
$\Theta$.

When $\alpha=\theta_{T}$ the inequality in (\ref{ineg}) turns to equality,
which yields
\begin{equation}
\phi(\theta,\theta_{T})=\sup_{\alpha\in\mathcal{F}_{\theta}}\int
\varphi^{\prime}\left(  \frac{dP_{\theta}}{dP_{\alpha}}\right)  dP_{\theta
}-\int\varphi^{\#}\left(  \frac{dP_{\theta}}{dP_{\alpha}}\right)
dP_{\theta_{T}} \label{forme duale div}%
\end{equation}
Denote
\begin{equation}
h(\theta,\alpha,x):=\int\varphi^{\prime}\left(  \frac{dP_{\theta}}{dP_{\alpha
}}\right)  dP_{\theta}-\varphi^{\#}\left(  \frac{dP_{\theta}}{dP_{\alpha}%
}\right)  \label{functionh}%
\end{equation}
from which
\begin{equation}
\phi(\theta,\theta_{T})=\sup_{\alpha\in\mathcal{F}_{\theta}}\int
h(\theta,\alpha,x)dP_{\theta_{T}}. \label{forme dual div pour est}%
\end{equation}
Furthermore by (\ref{ineg}), for all suitable $\alpha$
\begin{align*}
&  \phi(\theta,\theta_{T})-\int h(\theta,\alpha,x)dP_{\theta_{T}}\\
&  =\int h(\theta,\theta_{T},x)dP_{\theta_{T}}.-\int h(\theta,\alpha
,x)dP_{\theta_{T}}\geq0
\end{align*}
and the function $x\rightarrow h(\theta,\theta_{T},x)-h(\theta,\alpha,x)$ is
non negative, due to (\ref{convevity of phi}). It follows that $\phi
(\theta,\theta_{T})-\int h(\theta,\alpha,x)dP_{\theta_{T}}$ is zero only if
$h(\theta,\alpha,x)=h(\theta,\theta_{T},x)-$ $P_{\theta_{T}}$ a.e. Therefore
for any $x$ in the support of $P_{\theta_{T}}$
\[
\left[  \int\varphi^{\prime}\left(  \frac{dP_{\theta}}{dP_{\theta_{T}}%
}\right)  dP_{\theta}-\int\varphi^{\prime}\left(  \frac{dP_{\theta}%
}{dP_{\alpha}}\right)  dP_{\theta}\right]  -\varphi^{\#}\left(  \frac
{dP_{\theta}}{dP_{\alpha}}(x)\right)  +\varphi^{\#}\left(  \frac{dP_{\theta}%
}{dP_{\theta_{T}}}\left(  x\right)  \right)  =0
\]
which cannot hold for all $x$ when the functions $\varphi^{\#}\left(
\frac{dP_{\theta}}{dP_{\alpha}}(x)\right)  ,$ $\varphi^{\#}\left(
\frac{dP_{\theta}}{dP_{\theta_{T}}}\left(  x\right)  \right)  $ and $1$ are
linearly independent, unless $\alpha=\theta_{T}.$ We have proved that
$\theta_{T}$ is the unique optimizer in (\ref{forme duale div}).

We have skipped some sufficient conditions which ensure that (\ref{ineg}) holds.

Assume that
\begin{equation}
\int\left\vert \varphi^{\prime}\left(  \frac{dP_{\theta}}{dP_{\alpha}}\right)
\right\vert dP_{\theta}<\infty. \label{p_teta/p_alfain L1(teta)}%
\end{equation}
Assume further that $\phi(\theta,\theta_{T})$ is finite. Since
\begin{align*}
-\int\varphi^{\#}\left(  \frac{dP_{\theta}}{dP_{\alpha}}(x)\right)
dP_{\theta_{T}}  &  \leq\phi(\theta,\theta_{T})-\int\varphi^{\prime}\left(
\frac{dP_{\theta}}{dP_{\theta_{T}}}\right)  dP_{\theta}\\
&  \leq\phi(\theta,\theta_{T})+\int\left\vert \varphi^{\prime}\left(
\frac{dP_{\theta}}{dP_{\theta_{T}}}\right)  \right\vert dP_{\theta}<+\infty
\end{align*}
we obtain%
\[
\int\varphi^{\#}\left(  \frac{dP_{\theta}}{dP_{\alpha}}\right)  dP_{\theta
_{T}}>-\infty
\]
which entails (\ref{ineg}). When $\int\varphi^{\#}\left(  \frac{dP_{\theta}%
}{dP_{\alpha}}\right)  dP_{\theta_{T}}=+\infty$ then clearly , under
(\ref{p_teta/p_alfain L1(teta)})
\[
\phi(\theta,\theta_{T})>\int\varphi^{\prime}\left(  \frac{dP_{\theta}%
}{dP_{\alpha}}\right)  dP_{\theta}-\int\varphi^{\#}\left(  \frac{dP_{\theta}%
}{dP_{\alpha}}\right)  dP_{\theta_{T}}=-\infty.
\]
We have proved that (\ref{forme duale div}) holds when $\alpha$ satisfies
(\ref{p_teta/p_alfain L1(teta)}).

Sufficient and simple conditions encompassing (\ref{p_teta/p_alfain L1(teta)})
can be assessed under standard requirements for nearly all divergences. We
state the following Lemma (see Liese and Vajda (1987)\cite{LV1987}) and
Broniatowski and K\'{e}ziou (2006) \cite{BK2006}, Lemma 3.2).

\begin{lemma}
Assume that \textbf{RC} holds and $\phi(\theta,\alpha)$ is finite. Then
(\ref{p_teta/p_alfain L1(teta)}) holds.
\end{lemma}

Summing up, \ we state

\begin{theorem}
Let $\theta$ belong to $\Theta$ and let $\phi(\theta,\theta_{T})$ be finite.
Assume that \textbf{RC} holds.Let $\mathcal{F}_{\theta}$ be the subset of all
$\alpha$'s in $\Theta$ such that $\phi(\theta,\alpha)$ is finite . Then
\[
\phi(\theta,\theta_{T})=\sup_{\alpha\in\mathcal{F}_{\theta}}\int
\varphi^{\prime}\left(  \frac{dP_{\theta}}{dP_{\alpha}}\right)  dP_{\theta
}-\int\varphi^{\#}\left(  \frac{dP_{\theta}}{dP_{\alpha}}\right)
dP_{\theta_{T}}.
\]
Furthermore the sup is reached at $\theta_{T}$ and uniqueness holds.
\end{theorem}

For the Cressie-Read family of divergences with $\gamma\neq0,1$ this
representation writes%
\[
\phi_{\gamma}(\theta,\theta_{T})=\sup_{\alpha\in\mathcal{F}_{\theta}}\left\{
\frac{1}{\gamma-1}\int\left(  \frac{dP_{\theta}}{dP_{\alpha}}\right)
^{\gamma-1}~dP_{\theta}-\frac{1}{\gamma}\int\left(  \frac{dP_{\theta}%
}{dP_{\alpha}}\right)  ^{\gamma}~dP_{\theta_{T}}-\frac{1}{\gamma(\gamma
-1)}\right\}  .
\]
The set $\mathcal{F}_{\theta}$ may depend on the choice of the parameter
$\theta$. Such is the case for the $\chi^{2}$ divergence i.e. $\varphi
(x)=(x-1)^{2}/2,$ when $p_{\theta}(x)=\theta\exp(-\theta x){\Large 1}_{\left[
0,\infty\right)  }(x).$ In most cases the difficulty of dealing with a
specific set $\mathcal{F}_{\theta}$ depending on $\theta$ can be encompassed
when
\begin{align}
&  \text{There exists a neighborhood }\mathcal{U}\text{ of }\theta_{T}\text{
for which}\tag{A}\label{(a)}\\
&  \text{ }\phi(\theta,\theta^{\prime})\text{ is finite whatever }\theta\text{
and }\theta^{\prime\text{ \ }}\text{in \ }\mathcal{U}\nonumber
\end{align}
which for example holds in the above case for any $\theta_{T}.$ This
simplication deserves to be stated in the next result

\begin{theorem}
When $\phi(\theta,\theta_{T})$ is finite and \textbf{RC} holds, then under
condition (A)
\[
\phi(\theta,\theta_{T})=\sup_{\alpha\in\mathcal{U}}\int\varphi^{\prime}\left(
\frac{dP_{\theta}}{dP_{\alpha}}\right)  dP_{\theta}-\int\varphi^{\#}\left(
\frac{dP_{\theta}}{dP_{\alpha}}\right)  dP_{\theta_{T}}.
\]
Furthermore the sup is reached at $\theta_{T}$ and uniqueness holds.
\end{theorem}

\begin{remark}
Identifying $\mathcal{F}_{\theta}$ might be cumbersome. This difficulty also
appears in the classical MLE case, a special case of the above statement with
divergence function $\varphi_{0}$ $,$for which it is assumed that
\[
\int\log p_{\theta}(x)p_{\theta_{T}}(x)d\lambda(x)\text{ is finite}%
\]
for $\theta$ in a neighborhood of $\theta_{T}.$
\end{remark}

Under the above notation and hypotheses define
\begin{equation}
T_{\theta}\left(  P_{\theta_{T}}\right)  :=\arg\sup_{\alpha\in\mathcal{F}%
_{\theta}}\int h(\theta,\alpha,x)dP_{\theta_{T}}. \label{FisherDefDFiEst}%
\end{equation}
It then holds%
\[
T_{\theta}\left(  P_{\theta_{T}}\right)  =\theta_{T}%
\]
for all $\theta_{T}$ in $\Theta$. Also let
\begin{equation}
S\left(  P_{\theta_{T}}\right)  :=\arg\inf_{\theta\in\Theta}\sup_{\alpha
\in\mathcal{F}_{\theta}}\int h(\theta,\alpha,x)dP_{\theta_{T}}.
\label{FisherDefMDFiEst}%
\end{equation}
which also satisfies%
\[
S\left(  P_{\theta_{T}}\right)  =\theta_{T}%
\]
for all $\theta_{T}$ in $\Theta$. We thus state

\begin{theorem}
When $\phi(\theta,\theta_{T})$ is finite for all $\theta$ in $\Theta$ and
\textbf{RC} holds, both functionals $T_{\theta}$ and $S$ are Fisher consistent
for all $\theta_{T}$ in $\Theta.$
\end{theorem}

\section{\bigskip Plug in estimators}

From (\ref{forme dual div pour est}) simple estimators for $\theta_{T}$ can be
defined, plugging any convergent empirical measure in place of $P_{\theta_{T}%
}$ and taking the infimum in $\theta$ in the resulting estimator of
$\phi(\theta,\theta_{T}).$

\bigskip In the context of simple i.i.d. sampling, introducing the empirical
measure
\[
P_{n}:=\frac{1}{n}\sum_{i=1}^{n}\delta_{X_{i}}%
\]
\bigskip where the $X_{i}$'s are i.i.d. r.v's with common unknown distribution
$P_{\theta_{T}}$ in $\mathcal{P},$ the natural estimator of $\phi
(\theta,\theta_{T})$ is
\begin{align}
\phi_{n}(\theta,\theta_{T})  &  :=\sup_{\alpha\in\mathcal{F}_{\theta}}\left\{
\int h(\theta,\alpha,x)~dP_{n}(x)\right\} \label{estimateur de phi}\\
&  =\sup_{\alpha\in\mathcal{F}_{\theta}}\int\varphi^{\prime}\left(
\frac{dP_{\theta}}{dP_{\alpha}}\right)  dP_{\theta}-\frac{1}{n}\sum_{i=1}%
^{n}\varphi^{\#}\left(  \frac{dP_{\theta}}{dP_{\alpha}}\left(  X_{i}\right)
\right)  .\nonumber
\end{align}

\noindent Since
\[
\inf_{\theta\in\Theta}\phi(\theta,\theta_{T})=\phi(\theta_{T},\theta_{T})=0
\]
the resulting estimator of $\phi(\theta_{T},\theta_{T})$ is
\begin{equation}
\phi_{n}(\theta_{T},\theta_{T}):=\inf_{\theta\in\Theta}\phi_{n}(\theta
,\theta_{T})=\inf_{\theta\in\Theta}\sup_{\alpha\in\mathcal{F}_{\theta}%
}\left\{  \int h(\theta,\alpha,x)~dP_{n}(x)\right\}  . \label{def Div Estim}%
\end{equation}
Also the estimator of $\theta_{T}$ is obtained as

$\bigskip$%
\begin{equation}
\widehat{\theta}:=\arg\inf_{\theta\in\Theta}\sup_{\alpha\in\mathcal{F}%
_{\theta}}\left\{  \int h(\theta,\alpha,x)~dP_{n}(x)\right\}  .
\label{def EMphiD estimates}%
\end{equation}
When \ref{(a)} holds then $\mathcal{F}_{\theta}$ may be substituted by
$\mathcal{U}$ in the above definitions.

The resulting minimum dual divergence estimators (\ref{def Div Estim}) and
(\ref{def EMphiD estimates}) do not require any smoothing or grouping, in
contrast with the classical approach which involves quantization.\ The paper
\cite{BK2009} provides a complete study of those estimates and subsequent
inference tools for the usual i.i.d. sample scheme$.$ For all divergences
considered here, these estimators are asymptotically efficient in the sense
that they achieve the Cramer-Rao bound asymptotically. The case when
$\varphi=\varphi_{0}$ leads to $\theta_{ML}$ defined as the celebrated Maximum
Likelihood Estimator (MLE), in the context of the simple sampling.

\section{Minimum divergence estimators in exponential families}

In this section we prove the following result

\begin{theorem}
For all divergence $\phi$ defined through a differentiable function $\varphi$
satisfying Condition (\textbf{RC}), the minimum dual divergence estimator
defined by (\ref{def EMphiD estimates}) coincides with the MLE on any full
exponential families such that $\phi\left(  \theta,\alpha\right)  $ is finite
for all $\theta$ and $\alpha$ in $\Theta.$
\end{theorem}

Let $\mathcal{P}$ be an exponential family on $\mathbb{R}^{s}$ with canonical
parameter in $\mathbb{R}^{d}$
\[
\mathcal{P}:=\left\{
\begin{array}
[c]{c}%
P_{\theta}\text{ such that }p_{\theta}(x)=\frac{dP_{\theta}}{d\lambda}(x)\\
=\exp\left[  T(x)^{\prime}\theta-C(\theta)\right]  \text{; }\theta\in\Theta
\end{array}
\right\}
\]
where $x$ is in $\mathbb{R}^{s}$ and $\Theta$ is an open subset of
$\mathbb{R}^{d}$ , and $\lambda$ is a dominating measure for $\mathcal{P}.$ We
assume $\mathcal{P}$ to be full, namely that the Hessian matrix $\left(
\partial^{2}/\partial\theta^{2}\right)  C(\theta)$ is definite positive for
all $\theta$ in $\Theta.$

Let $X_{1},...,X_{n}$ be $n$ \ i.i.d. random variables with common
distribution $P_{\theta_{T}\text{ }}$ with $\theta_{T}$ in $\Theta.$
Introduce
\[
M_{n}\left(  \theta,\alpha\right)  :=\int\varphi^{\prime}\left(
\frac{dP_{\theta}}{dP_{\alpha}}\right)  dP_{\theta}-\frac{1}{n}\sum_{i=1}%
^{n}\varphi^{\#}\left(  \frac{dP_{\theta}}{dP_{\alpha}}\left(  X_{i}\right)
\right)
\]

We will prove that
\begin{equation}
\inf_{\theta}\sup_{\alpha}M_{n}\left(  \theta,\alpha\right)  =0
\label{infsupM_n=0}%
\end{equation}
whatever the function $\varphi$ satisfying the claim. In (\ref{infsupM_n=0})
$\theta$ and $\alpha$ run in $\Theta.$ This result extends the maximum
likelihood case for which $\inf_{\theta}\sup_{\alpha}M_{n}\left(
\theta,\alpha\right)  =\sup_{\theta}\inf_{\alpha}\left[  \frac{1}{n}\sum
_{i=1}^{n}\log p_{\theta}\left(  X_{i}\right)  -\frac{1}{n}\sum_{i=1}^{n}\log
p_{\alpha}\left(  X_{i}\right)  \right]  =0.$

Direct substitution shows that for any $\theta,$
\[
\sup_{\alpha}M_{n}\left(  \theta,\alpha\right)  \geq M_{n}\left(
\theta,\theta\right)  =0
\]
from which
\begin{equation}
\inf_{\theta}\sup_{\alpha}M_{n}\left(  \theta,\alpha\right)  \geq0
\label{ineg sup}%
\end{equation}

\bigskip We prove that
\begin{equation}
\alpha=\theta_{ML}\text{ is the unique maximizer of }M_{n}\left(  \theta
_{ML},\alpha\right)  \label{alfaMaximise}%
\end{equation}
which yields
\begin{equation}
\inf_{\theta}\sup_{\alpha}M_{n}\left(  \theta,\alpha\right)  \leq\sup_{\alpha
}M_{n}\left(  \theta_{ML},\alpha\right)  =M_{n}\left(  \theta_{ML},\theta
_{ML}\right)  =0 \label{ineg inf}%
\end{equation}

which together with (\ref{ineg sup})\bigskip\ completes the proof.

Define
\begin{align*}
M_{n,1}\left(  \theta,\alpha\right)   &  :=\int\varphi^{\prime}\left(  \exp
A(\theta,\alpha,x)\right)  \exp B\left(  \theta,x\right)  d\lambda(x)\\
M_{n,2}\left(  \theta,\alpha\right)   &  :=\frac{1}{n}\sum_{i=1}^{n}%
\exp\left(  A\left(  \theta,\alpha,X_{i}\right)  \right)  \varphi^{\prime
}\left(  \exp A(\theta,\alpha,X_{i})\right) \\
M_{n,3}\left(  \theta,\alpha\right)   &  :=\frac{1}{n}\sum_{i=1}^{n}%
\varphi\left(  \exp A(\theta,\alpha,X_{i})\right)
\end{align*}
with
\begin{align*}
A(\theta,\alpha,x)  &  :=T(x)^{\prime}\left(  \theta-\alpha\right)
+C(\alpha)-C(\theta)\\
B(\theta,x)  &  :=T(x)^{\prime}\theta-C(\theta).
\end{align*}

It holds
\[
M_{n}\left(  \theta,\alpha\right)  =M_{n,1}\left(  \theta,\alpha\right)
-M_{n,2}\left(  \theta,\alpha\right)  +M_{n,3}\left(  \theta,\alpha\right)
\]
with
\[
\frac{\partial}{\partial\alpha}M_{n,1}\left(  \theta,\alpha\right)
_{\alpha=\theta}=-\varphi^{(2)}\left(  1\right)  \left[  \nabla C\left(
\theta\right)  -\nabla C\left(  \alpha\right)  _{\alpha=\theta}\right]  =0
\]
for all $\theta,$%
\[
\frac{\partial}{\partial\alpha}M_{n,2}\left(  \theta_{ML},\alpha\right)
_{\alpha=\theta_{ML}}=\varphi^{(2)}\left(  1\right)  \frac{1}{n}\sum_{i=1}%
^{n}\left[  -T(X_{i})+\nabla C\left(  \alpha\right)  _{\alpha=\theta_{ML}%
}\right]  =0
\]
and
\[
\frac{\partial}{\partial\alpha}M_{n,3}\left(  \theta_{ML},\alpha\right)
=\frac{1}{n}\sum_{i=1}^{n}\left[  -T(X_{i})+\nabla C\left(  \alpha\right)
_{\alpha=\theta_{ML}}\right]  =0
\]
where the two last displays hold iff $\alpha=\theta_{ML}.$ Now
\begin{align*}
\frac{\partial^{2}}{\partial\alpha^{2}}M_{n,1}\left(  \theta_{ML}%
,\alpha\right)  _{\alpha=\theta_{ML}}  &  =\left(  \varphi^{(3)}%
(1)+2\varphi^{(2)}(1)\right)  \left(  \partial^{2}/\partial\theta^{2}\right)
C(\theta_{ML})\\
\frac{\partial^{2}}{\partial\alpha^{2}}M_{n,2}\left(  \theta_{ML}%
,\alpha\right)  _{\alpha=\theta_{ML}}  &  =\left(  \varphi^{(3)}%
(1)+4\varphi^{(2)}(1)\right)  \left(  \partial^{2}/\partial\theta^{2}\right)
C(\theta_{ML})\\
\frac{\partial^{2}}{\partial\alpha^{2}}M_{n,3}\left(  \theta_{ML}%
,\alpha\right)  _{\alpha=\theta_{ML}}  &  =\varphi^{(2)}(1)\left(
\partial^{2}/\partial\theta^{2}\right)  C(\theta_{ML}),
\end{align*}
whence

\bigskip%
\begin{align*}
\frac{\partial}{\partial\alpha}M_{n}\left(  \theta_{ML},\alpha\right)
_{\alpha=\theta_{ML}}  &  =0\\
\frac{\partial^{2}}{\partial\alpha^{2}}M_{n}\left(  \theta_{ML},\alpha\right)
_{\alpha=\theta_{ML}}  &  =-\varphi^{(2)}(1)\left(  \partial^{2}%
/\partial\theta^{2}\right)  C(\theta_{ML})
\end{align*}
which proves (\ref{alfaMaximise}), and closes the proof.

\end{document}